\documentclass{IEEEtran}

\usepackage{url}
\usepackage{graphicx}
\usepackage{color}
\usepackage{placeins}
\usepackage{float}
\usepackage{hyperref}
\usepackage{tabularx,colortbl}
\usepackage{ifthen}
\usepackage{cite}
\usepackage{amsmath,amsfonts}
\usepackage{mathtools}
\usepackage{algorithmic}
\usepackage{textcomp}
\usepackage{xcolor}
\usepackage{comment}
\usepackage{siunitx}
\usepackage{pgfplots}
\pgfplotsset{compat=1.18}
\usetikzlibrary{pgfplots.colorbrewer}
\usepackage{tikz}
\usepackage{caption}
\usepackage{subcaption}
\usepackage{dsfont}
\usepackage{lipsum}
\usepackage{soul}
\usepackage{stackrel}

\usepackage{pgfkeys}
\newenvironment{customlegend}[1][]{%
    \begingroup
    \csname pgfplots@init@cleared@structures\endcsname
    \pgfplotsset{#1}%
}{%
    \csname pgfplots@createlegend\endcsname
    \endgroup
}%
\def\addlegendimage{\csname pgfplots@addlegendimage\endcsname}

\DeclareMathAlphabet\mathbfcal{OMS}{cmsy}{b}{n}
\DeclareMathOperator{\curl}{curl}
\DeclareMathOperator{\grad}{grad}
\DeclareMathOperator{\Div}{div}

\newcommand{\scalar}[1]{{\left\langle #1 \right\rangle}}
\newcommand{\dd}{\,\mathrm{d}}
\newcommand{\trans}{^{\!\top}}

\ifCLASSINFOpdf

\else

\fi

\usepackage{eso-pic}
\AddToShipoutPicture*{\footnotesize\sffamily\raisebox{0.75cm}{\hspace{1.65cm}\fbox{\parbox{\textwidth}{\copyright 2024 IEEE.  Personal use of this material is permitted.  Permission from IEEE must be obtained for all other uses, in any current or future media, including reprinting/republishing this material for advertising or promotional purposes, creating new collective works, for resale or redistribution to servers or lists, or reuse of any copyrighted component of this work in other works.}}}}

\begin{document}

\title{A Low-Frequency-Stable\\Higher-Order Isogeometric Discretization of the\\Augmented Electric Field Integral Equation}

\author{Maximilian~Nolte,
        Riccardo~Torchio,
        Sebastian~Schöps,
        Jürgen~Dölz,
        Felix~Wolf
        and~Albert~E.~Ruehli,~\IEEEmembership{Life~Fellow,~IEEE}
\thanks{%
Maximilian Nolte, Sebastian Schöps and Felix Wolf are with the Computational Electromagnetics Group, Technische Universität Darmstadt, Germany.
(Corresponding author: Maximilian Nolte, maximilian.nolte@tu-darmstadt.de.)}%
\thanks{Riccardo Torchio is with the Department of Industrial Engineering and the Department of Information Engineering of the
Università degli Studi di Padova, Italy.
}%
\thanks{%
Jürgen Dölz is with the Institute for Numerical Simulation, University of Bonn, Germany.
}%
\thanks{%
Albert E. Ruehli is with the EMC Laboratory, Missouri University of
Science and Technology, Rolla, MO USA.}%
}

\maketitle

\begin{abstract} 
This contribution investigates the connection between isogeometric analysis and integral equation methods for full-wave electromagnetic problems up to the low-frequency limit.
The proposed spline-based integral equation method allows for an exact representation of the model geometry described in terms of non-uniform rational B-splines without meshing.
This is particularly useful when high accuracy is required or when meshing is cumbersome for instance during optimization of electric components.
The augmented electric field integral equation is adopted and the deflation method is applied, so the low-frequency breakdown is avoided.
The extension to higher-order basis functions is analyzed and the convergence rate is discussed.
Numerical experiments on academic and realistic test cases demonstrate the high accuracy of the proposed approach.
\end{abstract}

\begin{IEEEkeywords}
integral equations (IEs), computer aided design (CAD), B-splines, isogeometric analysis (IGA), electromagnetic modeling.
\end{IEEEkeywords}

\section{Introduction}
Accurate modeling of electromagnetic components is paramount for optimizing electric and electronic devices.
Among all numerical methods, integral equation (IE) methods have been widely and effectively utilized for, e.g., scattering, radiation and optimal control problems~\cite{Jin_2010aa_MoM,Colton_2013aa_Ch9,Adrian_2021aa}.
A multitude of formulations have been devised to tackle the electromagnetic challenges that emerge from various engineering contexts.

Many attempts have been made to enhance the accuracy and efficiency of IE methods across wide frequency ranges, i.e., from direct current (dc) to very high frequencies.  
In particular, techniques have been proposed to solve the low-frequency breakdown of the electric field integral equation (EFIE), which is known from the early '80s \cite{Wilton_1981aa,Mautz_1984aa}.
For instance, the loop-star decomposition~\cite{Zhao_2000aa,Andriulli_2012aa,Bourhis_2024aa} and current-charge integral equations~\cite{Taskinen_2006aa} have been successfully incorporated to overcome the low-frequency breakdown (see, e.g., \cite{Qian_2010ab} for a list of references on this topic).
Another effective approach to address the low-frequency breakdown is the use of the augmented electric field integral equation (A-EFIE)~\cite{Qian_2008aa}.
This is a mixed formulation based on currents and charges or electric scalar potentials~\cite{Qian_2009aa} which resembles the formulation of the partial element equivalent circuit (PEEC) method~\cite{Ruehli_1974aa}.
Suitable scaling and enforcing charge neutrality guarantees low-frequency stability in the limit, see e.g. \cite{Mautz_1984aa,Taskinen_2006aa,Qian_2009aa,Sharma_2022aa}.

Commonly, implementations for IE methods rely on surface or volume meshes that approximate geometries using lowest-order elements, such as triangles and/or tetrahedra together with lowest-order basis functions to represent the solution~\cite{Xia_2016aa,Zhang_2022ab}.
However, this approach comes with several notable drawbacks.
As such an example, entirely lowest-order approaches require many degrees of freedom to obtain high accuracy and higher-order basis functions may not unlock their full potential if the geometrical error hinders the overall convergence of the method, see~\cite[Sec.~4]{Strang_2008aa} for details.
Moreover, the meshing of geometries from computer aided design (CAD) can be time-consuming and prone to errors.
Its contribution to the overall workflow is significant; according to Sandia labs, approximately 75\% of simulation time is attributed to modeling, parameterization, mesh generation, and pre- and post-processing \cite{Boggs_2005aa}.

The first point has been addressed by research on higher-order basis functions, see e.g., \cite{Kang_2001aa,Graglia_1997aa} and references therein.
Together with this development, the curvature of geometries has gained interest as a research topic: 
\cite{Jorgensen_2004aa} proposes a higher-order hierarchical Legendre basis on polynomially curved surfaces while specifically addressing ill-conditioning.
Their numerical examples address basis functions up to order seven and second order curved geometry.
In \cite{Ganesh_2008aa}, a different higher-order basis is introduced for geometries given in terms of spherical harmonics.
A higher-order formulation based on a current-charge mixed problem is considered in~\cite[Sec.~5.3]{Weggler_2011aa}. More recently, the direct usage of CAD data has been addressed.
As an example, \cite{Bruno_2020aa,Hellicar_2008aa,Hu_2021aa} use higher-order (spectral) basis functions on CAD geometries given in terms non-uniform rational B-splines (NURBS).
A particular successful variant are isogeometric boundary element methods. They use splines not only for geometry representation but also as higher-order basis functions such that refinement and smoothness control become straightforwardly available.
Applications of this methodology have been demonstrated for electromagnetism, e.g., \cite{Li_2016aa,Simpson_2018aa,Dolz_2018aa,Dolz_2020aa,Fays_2023aa} and neighboring disciplines, e.g., acoustics \cite{Simpson_2014ac,Chen_2020aa}.
The Galerkin discretization of the isogeoemtric EFIE has been mathematically analyzed in~\cite{Dolz_2019ad}.
However, the formulation used is not stable up to the low-frequency limit.
This impasse has recently been resolved based on the loop-star decomposition~\cite{Hofmann_2024aa}.
To the best of the authors' knowledge, there is no publication on higher-order discretizations of the A-EFIE with potentials as unknowns which in addition guarantees low-frequency stability.

In this paper, a higher-order isogeometric A-EFIE formulation is proposed for the first time. It provides stability up to the low-frequency limit.
The main contribution is to derive the weak form that enables a stuitable higher-order discretization within the de Rham sequence~\cite{Buffa_2019ac}.
For stability down to the low-frequency limit, no basis modifications, e.g., known from quasi Helmholtz decompositions~\cite{Bourhis_2024aa,Hofmann_2024aa}, is required. This eases the implementation which is based here on
the open-source code Bembel \cite{Dolz_2020ac,Nolte_2024ab} and its implementation for electrostatics \cite{Torchio_2023aa}.

The remainder of the paper is structured as follows.
In Section~\ref{sec:SIE}, we state the problem formulation in a continuous form and introduce the Galerkin variational formulation.
The discretization with B-splines and NURBS is described in Section~\ref{sec::discretization}.
The numerical experiments in Section~\ref{sec:results} test the implementation with an academic and an application example and show high accuracy of the proposed method.
Finally, we conclude in Section~\ref{sec:conclusions}.
\section{Integral Equation Method} \label{sec:SIE}

In the following, we introduce a PEEC-like A-EFIE formulation based on \cite{Qian_2009aa,Gope_2007aa},  which is also suitable for discretizations with higher-order B-splines within an isogeometric setting~\cite{Hughes_2005aa}.

\subsection{Formulation}
\begin{figure}[!t]
    \centering
\includegraphics[width=0.5\textwidth]{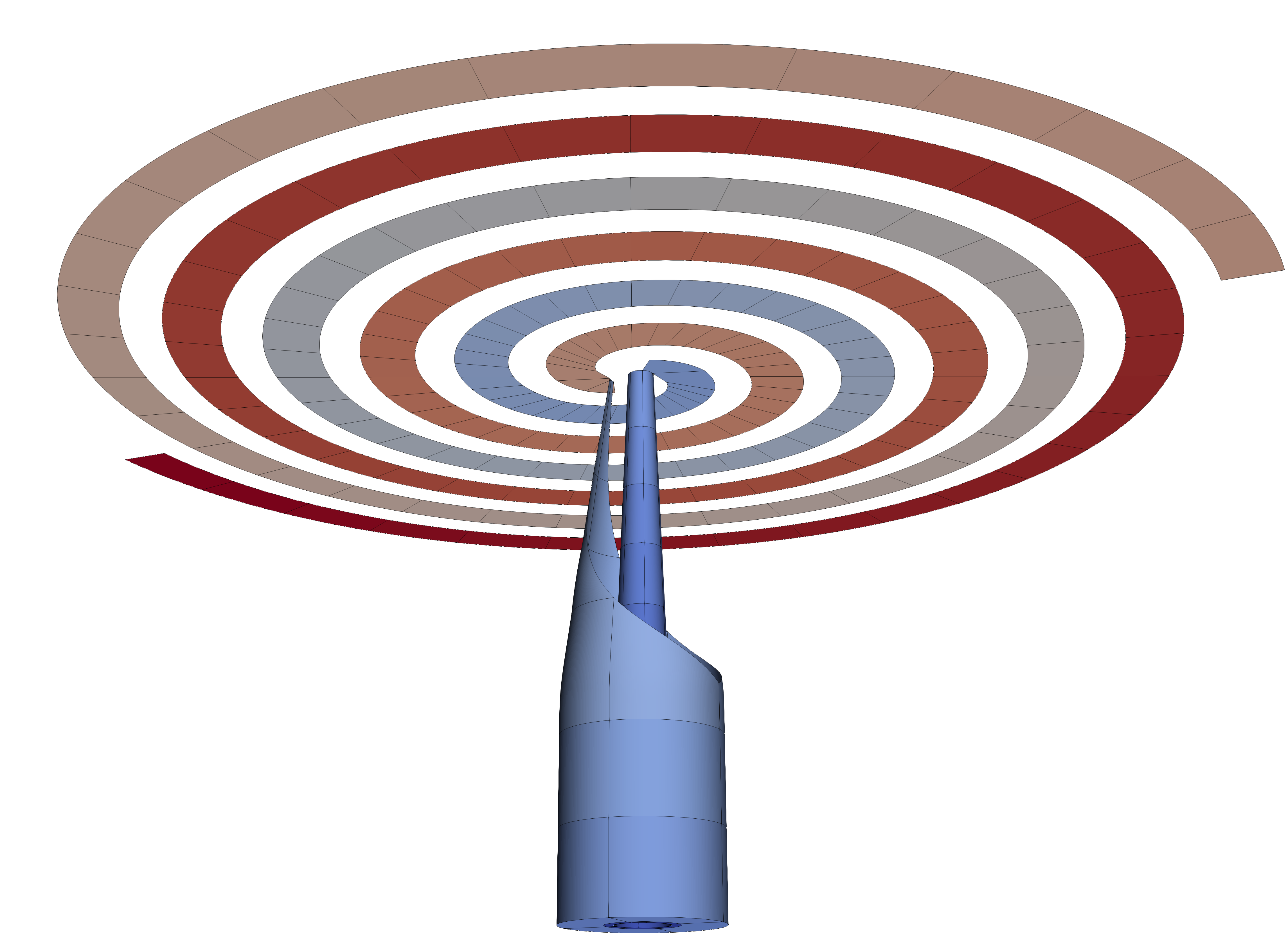}
\includegraphics[width=0.5\textwidth]{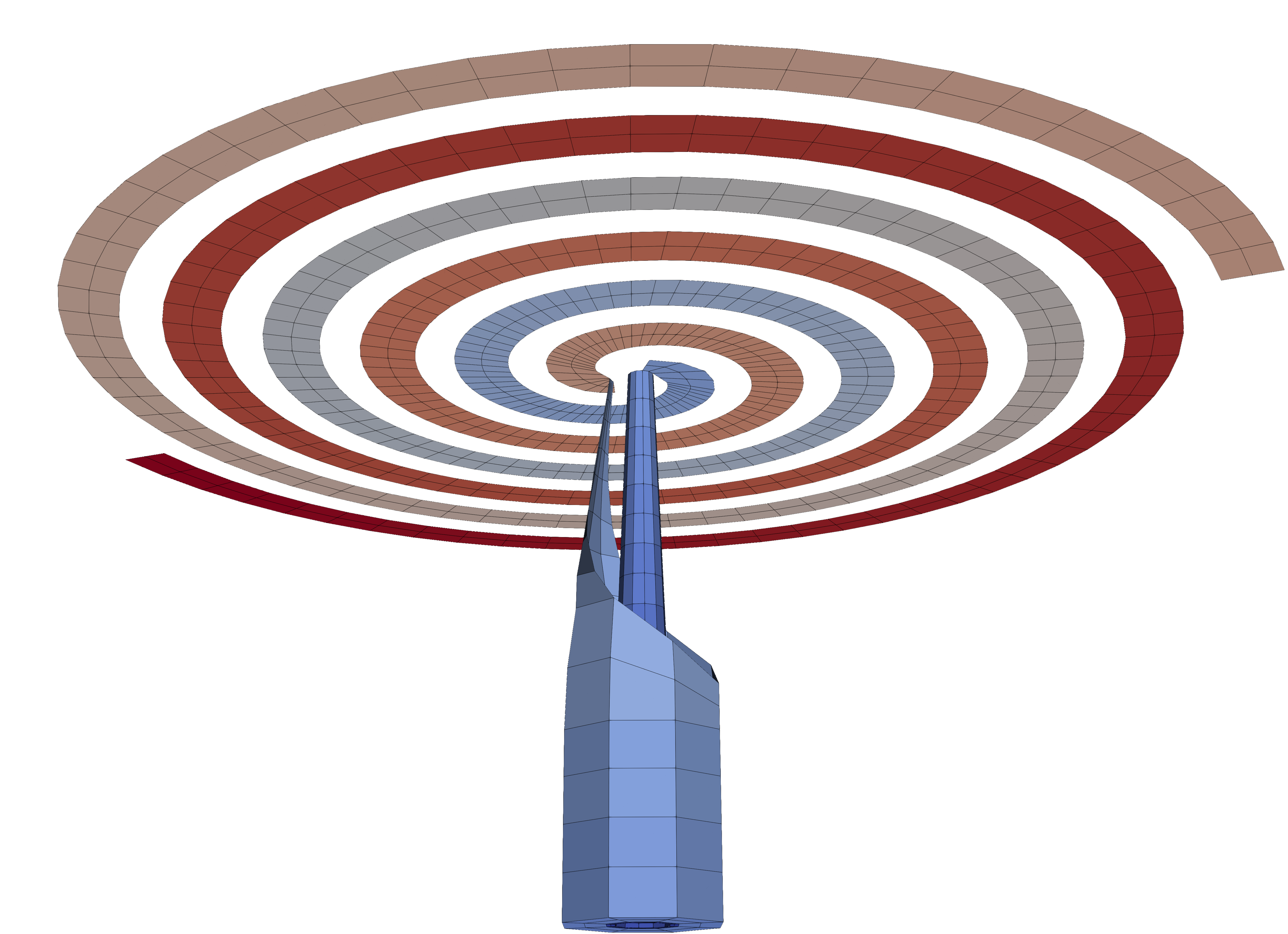}
    \caption{Visualization of the discretized PEC surface $\Gamma$ of a balun with spiral antenna.
    The discretization with $271$ NURBS patches (top) features exact geometry representation of the design proposed in \cite{McParland_2022aa}.
    For comparison, there is a conventional geometry approximation with $1084$ quadrilaterals~(bottom).}
\label{fig::balun_spiral_discretization}
\end{figure}
Let the geometry of a perfect electric conductor (PEC) surface be given as $\Gamma$, see for example Fig.~\ref{fig::balun_spiral_discretization}.
We follow \cite{Buffa_2003ab,Dolz_2019ad}, and assume Lipschitz continuity.
Furthermore, let $\Gamma$ be surrounded by homogeneous, linear and isotropic free-space $\Omega$ with permittivity~$\varepsilon$ and permeability~$\mu$.
For a given incident field~$\pmb{E}_{\mathrm{i}}$, where the harmonic time dependency $e^{j\omega t}$ is assumed and omitted in the following, we compute the scattered field~$\pmb{E}_{\mathrm{s}}$.
Thereby, if~$\pmb{n}_{\pmb{x}}$ denotes the outward normal vector on $\Gamma$ at point $\pmb{x}$, the following must hold
\begin{align}
    \pmb{E}_{\mathrm{s}}\times\pmb{n}_{\pmb{x}} &= -\pmb{E}_{\mathrm{i}} \times\pmb{n}_{\pmb{x}}, \quad\pmb{x}\in\Gamma.\label{eq::tangentialEfield}
\end{align}
The scattered field can be described using the well-known Stratton-Chu representation formular, cf.~\cite[Sec.~8.14]{Stratton_1941aa},
\begin{align}
    \pmb{E}_{\mathrm{s}}(\pmb x) = -j\omega \pmb{A}(\pmb{x}) - \pmb{\grad}\varphi(\pmb{x}),\quad\pmb{x}\in\Omega,\label{eq::EFIE}
\end{align}
where $\pmb{A}$ is the magnetic vector potential and $\varphi$ the electric scalar potential. These two potentials can be written as
\begin{alignat}{2}
    \pmb{A}(\pmb{x}) &&\coloneqq \pmb{A}[\pmb{j}](\pmb{x}) &= \mu\int_\Gamma \pmb{j}(\pmb{y})g_\kappa(\pmb{x},\pmb{y})\dd\Gamma_{\pmb{y}},\label{eq::BiotSavart}\\
    \varphi(\pmb{x}) &&\coloneqq C[\varrho](\pmb{x})  &= \frac{1}{\varepsilon}\int_\Gamma \varrho(\pmb y)g_\kappa(\pmb{x}, \pmb{y})\dd\Gamma_{\pmb{y}},\label{eq::CoulombIntegral}
\end{alignat}
with surface electric current $\pmb{j}$, electric charge $\varrho$, wave number~$\kappa =\omega\sqrt{\varepsilon\mu}$ and the homogeneous space Green's function 
\begin{equation}
g_\kappa(\pmb{x},\pmb{y}) = \frac{e^{-j\kappa|\pmb{x}-\pmb{y}|}}{4\pi|\pmb{x}-\pmb{y}|}.
\end{equation}
The surface $\Gamma$ can be either closed or open, taking care that the normal component of the current out of the surface at the boundary is zero \cite{Bendali_1984aa}.
For the mathematical analysis of the scattering problem solved with integral equations on screens the reader is referred to~\cite{Buffa_2003aa}.

Substituting the charge using the continuity equation
\begin{align}
    \Div_\Gamma\pmb{j} + j\omega\varrho = 0 \label{eq::continuity}
\end{align}
gives the EFIE in terms of surface electric current~$\pmb j$ which can be solved with the method of moments~\cite{Harrington_1968aa} using the well-known Rao-Wilton-Glisson basis functions~\cite{Rao_1982aa}.
Existence and uniqueness of the corresponding Galerkin solution on Lipschitz boundaries is proven in \cite{Buffa_2003ab}.

However, this formulation suffers from stability problems in the low-frequency limit~\cite{Mautz_1984aa,Zhao_2000aa}.
Alternative methods as presented in~\cite{Qian_2009aa,Xia_2016aa} state a mixed problem in terms of currents and charges, which is a similar approach to the PEEC method~\cite{Ruehli_1974aa,Ruehli_2015aa}.

In the next section we derive the variational formulation for the mixed problem in terms of currents and potentials as presented for the lowest-order case in~\cite{Qian_2009aa,Gope_2007aa}.
Stability issues of this formulation for the lossy case have been addressed in~\cite{Gope_2007aa}.
This formulation allows automatically for \textit{dc} solutions and it remains stable even in situations where there is no unique \textit{dc} solutions, e.g., for excitations that are inconsistent in the low-frequency limit.
This is important for example in many electromagnetic compatibility applications, see~\cite[Sec.~11.1]{Paul_2006aa}.
\subsection{Variational Formulation}
Before stating the variational formulation, we first fix notation of the required function spaces and duality pairings for the Galerkin method.
Their rigorous definition is beyond the scope of this contribution, so we refer to \cite{Buffa_2003ab},\cite{Bendali_1984aa} and \cite{Dolz_2019ad} for more details. 
Scalar quantities on the surface are described within Sobolev spaces $H^{1/2}(\Gamma)$ and its dual space $H^{-1/2}(\Gamma)$ with respect to the scalar product
\begin{align}
    \scalar{p, q} = \int_\Gamma p\overline{q}\dd\Gamma,\label{eq::L2scalarProduct}
\end{align}
with $p\in H^{1/2}(\Gamma)$ and $q\in H^{-1/2}(\Gamma)$ or vice versa.
The appropriate space to discretize the surface current density is~$\pmb{H}^{-1/2}_{\times}(\Div_\Gamma,\Gamma)$, where the reader is referred to~\cite{Buffa_2002aa} for a rigorous definition, with the corresponding duality pairing
\begin{align}
    \scalar{\pmb u,\pmb v}_\times = \int_\Gamma (\pmb u \times \pmb{n}_{\pmb x})\cdot \pmb{v}\dd\Gamma_{\pmb x},\label{eq::dualityPairing}
\end{align}
with $\pmb{u},\pmb{v}\in \pmb{H}^{-1/2}_{\times}(\Div_\Gamma,\Gamma)$.
For the variational formulation of the mixed problem, the tangential components in \eqref{eq::tangentialEfield} with~$\eqref{eq::EFIE}$ inserted is tested with respect to~\eqref{eq::dualityPairing}.
Furthermore, as proposed in~\cite[Sec.~2.2]{Bendali_1984aa} we test \eqref{eq::continuity} with respect to \eqref{eq::L2scalarProduct} with $C[\varphi']$, in which $\varphi'\in H^{-1/2}(\Gamma)$.
The variational problem then reads:
For a given sufficiently smooth incident field $\pmb{E}_{\mathrm{i}}$ find \mbox{$(\pmb{j}, \varphi)\in \pmb{H}^{-1/2}_{\times}(\Div_\Gamma,\Gamma) \times H^{1/2}(\Gamma)$} such that
\begin{align}
    \label{eq::productOperators0}
    j\omega\scalar{\pmb{A}[\pmb{j}],\pmb{j}'}_\times + \scalar{\pmb{\grad}_\Gamma \varphi, \pmb{j}'}_\times &= \scalar{\pmb{E}_{\mathrm{i}},\pmb{j}'}_\times,\\
    \scalar{C[\Div_\Gamma \pmb{j}], \varphi'} + j\omega\scalar{\varphi, \varphi'} &= 0,\label{eq::productOperators}
\end{align}
for all $(\pmb{j}',\varphi')\in \pmb{H}^{-1/2}_{\times}(\Div_\Gamma,\Gamma)\times H^{-1/2}(\Gamma)$.
For the second equation, it is used that $C$ is self-adjoint, i.e., can be switched in the slots of \eqref{eq::L2scalarProduct} and then $\varphi = C[\varrho]$ is substituted which yields the formulation in terms of the current and potential.
\section{Discretization}\label{sec::discretization}
This section is concerned with the discretization of the mixed A-EFIE formulation \mbox{\eqref{eq::productOperators0}-\eqref{eq::productOperators}}.
First, we discuss consistent discretization spaces, quadrature, system matrix assembly and finally low-frequency stabilization.
\begin{figure}[t!]
    \centering
    \input{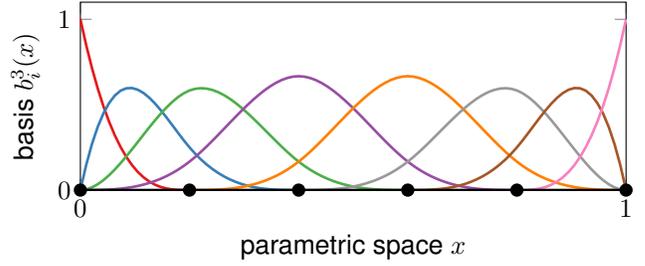}
    \caption{B-spline basis for $p=3$ with six control points.}
    \label{fig::bslines}
\end{figure}
\begin{figure*}
    \centering
    \input{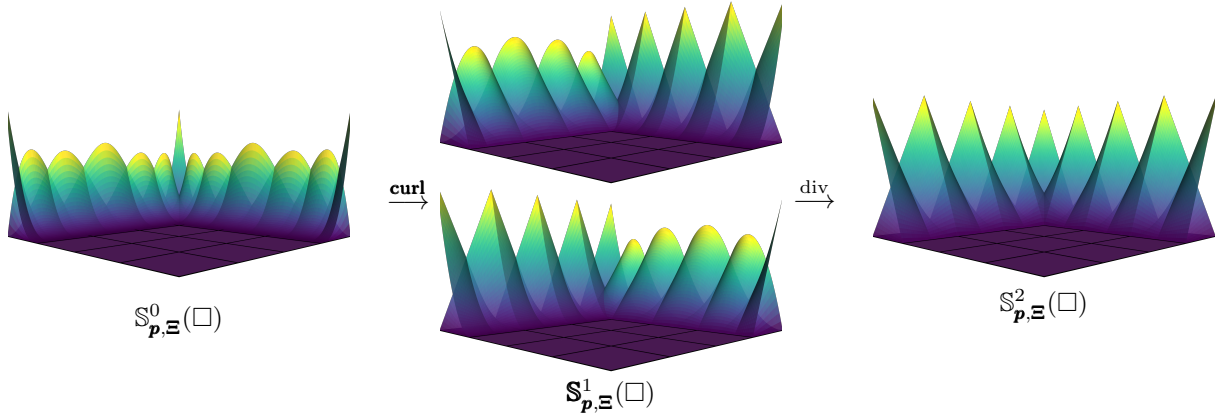}
    \caption{
    Illustration of the tensor product B-spline basis functions on one patch for $\pmb{p}=(2,2)$.
    To visualize the univariate B-spline basis related to each coordinate direction utilized in the \mbox{definition \eqref{eq::definition::S0}-\eqref{eq::definition::S2}}, only basis function with support on one edge of the patch are depicted.
    Applying the two-dimensional $\pmb{\curl}$ operator lowers for each vector component the polynomial degree with respect to one coordinate direction.
    Both coordinate directions are then lowered by applying the $\Div$ operator.
    }
    \label{fig::deRahm_complex}
\end{figure*}
\subsection{Spline Spaces and Geometry Mapping} \label{sec:splines}
Most common CAD kernels use boundary representation by NURBS \cite{Piegl_1997aa} to describe the computational domain.
The basis functions are B-splines which are defined using a $p$-open knot vector $\Xi = [\xi_1,\dots,\xi_{k + p + 1}]\in[0,1]^{k+p+1}$, where $k$ denotes the number of control points \cite{Beirao-da-Veiga_2014aa}.
We can then define the basis functions for $x\in(0,1)$ and $1\leq i\leq k$ as
\begin{align}
    b_i^0(x) &= \begin{cases}
        1, \quad \mathrm{if}\;\xi_i \leq x < \xi_{i+1},\\
        0, \quad \mathrm{otherwise},
    \end{cases}
\intertext{and for polynomial degree $p>0$ via the recursive relationship}
    b_i^p(x) &= \frac{x - \xi_i}{\xi_{i+p} - \xi_i} b_i^{p-1}(x) +
    \frac{\xi_{i+p+1} - x}{\xi_{i+p+1} - \xi_{i+1}} b_{i+1}^{p-1}(x),
\end{align}
where the B-splines $b_i^p$ span the one-dimensional space $S^p(\Xi)$, cf.~\cite{Hughes_2005aa}.
Fig.~\ref{fig::bslines} shows a visualization of the basis functions of degree $p=3$.
Fixing polynomial degrees $p^n_1$ and $p^n_2$ for each patch $\Gamma_n$ the NURBS mapping from the two-dimensional reference domain to the computational domain is given by
\begin{align}
    \!\!\Gamma_n(x, y) =\!\sum_{j_1=1}^{k^n_1}\sum_{j_2=1}^{k^n_2}\frac{\pmb{c}_{j_1,j_2} b_{j_1}^{p^n_1}(x) b_{j_2}^{p^n_2}(y) w_{j_1,j_2}}{ \sum_{i_1=1}^{k^n_1}\sum_{i_2=1}^{k^n_2} b_{i_1}^{p^n_1}(x) b_{i_2}^{p^n_2}(y) w_{i_1,i_2}},\!\!\label{eq::NURBS}
\end{align}
with control points $\pmb{c}_{j_1,j_2}\in\mathbb{R}^3$ and weights $w_{i_1,i_2}$.
In engineering applications the computational domain is often given by a multi-patch decomposition $\Gamma = \bigcup_{n=1}^{N_\Gamma}\Gamma_n$, where $N_\Gamma$ denotes the number of patches.

Independent of the choice of basis functions for the geometry representation, we define for the discretization of the surface current and potential tensor product B-spline spaces.
The two-dimensional B-spline space is denoted by $S^{p_1,p_2}(\Xi_1,\Xi_2)$, where $p_1,p_2$ and $\Xi_1,\Xi_2$ denote the polynomial degrees and knot vectors, respectively, employed in each individual coordinate direction of the tensor product construction.
Let $\pmb{p} = (p_1,p_2)$ denote a pair of polynomial degrees with their corresponding $p_1$- and $p_2$-open knot vectors $\pmb{\Xi} = (\Xi_1,\Xi_2)$.
Furthermore, let $\pmb{\Xi}_i'$ with $i=1,2$ denote a truncated knot vector with their first and last knot removed.
Then we can define the spline spaces on the unit square $\square = (0,1)^2$ as
\begin{align}
    \mathbb{S}^0_{\pmb{p},\pmb{\Xi}}(\square) 
    &\coloneqq
     S^{p_1,p_2}(\Xi_1,\Xi_2),\label{eq::definition::S0}
    \\
    \pmb{\mathbb{S}}^1_{\pmb{p},\pmb{\Xi}}(\square) 
    &\coloneqq
    S^{p_1,p_2-1}(\Xi_1,\Xi_2')\times S^{p_1-1,p_2}(\Xi_1',\Xi_2),\label{eq::definition::S1}
    \\
    \mathbb{S}^2_{\pmb{p},\pmb{\Xi}}(\square) 
    &\coloneqq
    S^{p_1-1,p_2-1}(\Xi_1',\Xi_2'),\label{eq::definition::S2}
\end{align}
which comply with the traces of the de Rham sequence, see~\cite{Dolz_2019ad}.
In Fig.~\ref{fig::deRahm_complex} the spline spaces for the case of quadratic basis functions are visualized.
The domain between two neighboring knots $\xi_i$ and $\xi_{i + 1}$ and its tensor product counter part are referred to as elements.

These B-spline spaces are defined on the reference domain~$\square$ and then mapped to the physical domain for each patch using~\eqref{eq::NURBS} and the corresponding Piola transformation, see e.g., \cite{Peterson_2006aa}.
A rigorous definition of the mapped function space $\pmb{\mathbb{S}}^1_{\pmb{p},\pmb{\Xi}}(\Gamma)\subset\pmb{H}^{-1/2}_{\times}(\Div_\Gamma,\Gamma)$, which provides normal continuity, can be found in~\cite{Buffa_2019ac}.
This allows the discretization of the surface current.
Furthermore, the potential is discretized in $\mathbb{S}^2_{\pmb{p},\pmb{\Xi}}(\Gamma)\subset H^{-1/2}(\Gamma)$, see also \cite{Buffa_2019ac} for a rigorous definition.
In the following we assume to have a suitable basis $\{\pmb{\nu}_1,\ldots,\pmb{\nu}_{N_{\pmb{j}}}\}$ of $\pmb{\mathbb{S}}^1_{\pmb{p},\pmb{\Xi}}(\Gamma)$ and $\{\varphi_1,\ldots,\varphi_{N_\varphi}\}$ of $\mathbb{S}^2_{\pmb{p},\pmb{\Xi}}(\Gamma)$ to be to our disposal.
For convenience, we denote the total number of degrees of freedom by $N = N_{\pmb{j}} + N_{\varphi}$.

\subsection{Assembly and Quadrature}
In the following, we describe the assembly of the system matrix and the right-hand side that arises when discretizing the variational problem \mbox{\eqref{eq::productOperators0}-\eqref{eq::productOperators}}.
Except for the bilinear form $\scalar{C[\Div_\Gamma \pmb{j}], \varphi'}$, all terms in the A-EFIE formulation can easily be discretized with higher-order basis functions and implementations are readily available in most software packages such as \cite{Dolz_2020ac}.
In the following we perform an additional approximation step which replaces the discrete version, i.e., matrix, of $\scalar{C[\Div_\Gamma \pmb{j}], \varphi'}$ by a product of matrices which are commonly implemented in IE software packages.

To this end, we observe that the operator $C$ is straightforwardly discretized with respect to the basis $\{\varphi_1,\ldots,\varphi_{N_\varphi}\}$ of~$\mathbb{S}^2_{\pmb{p},\pmb{\Xi}}(\Gamma)$, i.e., provided by the matrix $\pmb{P}$ with entries
\begin{equation}
    P_{ij} = \frac{1}{\varepsilon}\int_\Gamma \int_\Gamma g_\kappa(\pmb{x},\pmb{y}) \varphi_i \varphi_j \dd\Gamma_{\pmb{y}}\dd\Gamma_{\pmb{x}}.
\end{equation}
To incorporate the computation of $\Div_\Gamma\pmb{j}$ in terms of this basis we need to perform an appropriate ``change'' of basis.
Given the current is discretized by
\begin{equation}
    \pmb{j}^h=\sum_{i=1}^{N_{\pmb{j}}}J_i\pmb{\nu}_i,
\end{equation}
with $\{\pmb{\nu}_1,\ldots,\pmb{\nu}_{N_{\pmb{j}}}\}$ the basis of $\pmb{\mathbb{S}}^1_{\pmb{p},\pmb{\Xi}}(\Gamma)$ and degrees of freedom $\pmb{J}=(J_1,$ $\dots,$ $J_{N_{\pmb{j}}})\trans$, we are looking for a representation of $\Div_\Gamma \pmb{j}^h$ with respect to $\mathbb{S}^2_{\pmb{p},\pmb{\Xi}}(\Gamma)$, which we denote by 
\begin{equation}
    d^h=\sum_{i=1}^{N_\varphi}d_i\varphi_i.
\end{equation}
We achieve this approximately by solving the variational problem
\begin{align}
    \scalar{d^h,\varphi_i} = \scalar{\Div_\Gamma \pmb{j}^h,\varphi_i},
\end{align}
for all $\{\varphi_1,\ldots,\varphi_{N_\varphi}\}$.
Since it is finite dimensional, this is equivalent to solving the linear system
\begin{align}
    \pmb{Md} = \pmb{SJ},\label{eq::L2projection}
\end{align}
for \mbox{$\pmb{d}=(d_1,\dots d_{N_\varphi})\trans$} with matrix entries
\begin{align*}
    M_{ij} = \int_\Gamma \varphi_i \varphi_j \dd\Gamma_{\pmb{x}},\qquad S_{ij} &= \int_\Gamma \varphi_i \Div_\Gamma \pmb{\nu}_j \dd\Gamma_{\pmb{x}}.
\end{align*}
This allows to approximate the matrix of $\scalar{C[\Div_\Gamma \pmb{j}], \varphi'}$ by~$\pmb{P}\pmb{M}^{-1}\pmb{S}$.

In order to assemble the system matrix, we further need the discretization of $\varphi$ which is denoted by 
\begin{equation}
    \varphi^h=\sum_{i=1}^{N_\varphi}\Phi_i\varphi_i,
\end{equation}
with the degrees of freedom $\pmb{\Phi}=(\Phi_1,$ $\dots,$ $\Phi_{N_{\varphi}})\trans$.
Inserting $\pmb{j}^h$ and $\varphi^h$ into \mbox{\eqref{eq::productOperators0}-\eqref{eq::productOperators}} and testing with the basis $\{\pmb{\nu}_1,\ldots,\pmb{\nu}_{N_{\pmb{j}}}\}$ and $\{\varphi_1,\ldots,\varphi_{N_\varphi}\}$ yields
\begin{align}
  \begin{pmatrix}
     j\omega\pmb{L}\vphantom{\pmb{S}\trans} & \pmb{S}\trans\vphantom{j\omega\pmb{L}}
    \\ \pmb{P}\pmb{M}^{-1}\pmb{S} & -j\omega\pmb{M}
  \end{pmatrix}
  \begin{pmatrix}
    \pmb{J} \\ \pmb{\Phi}
  \end{pmatrix}
  =
  \begin{pmatrix}
    \pmb{v}_{\mathrm{ex}} \\ \pmb{0}
  \end{pmatrix}\label{eq::discritization::systemmatrix}.
\end{align}
The matrix entries for the inductive effects are computed as follows
\begin{equation}
    L_{ij} = \mu\int_\Gamma \int_\Gamma g_\kappa(\pmb{x},\pmb{y}) \pmb{\nu}_i \cdot \pmb{\nu}_j \dd\Gamma_{\pmb{y}}\dd\Gamma_{\pmb{x}}.
\end{equation}
The entries in the matrix blocks $\pmb{L}$ and $\pmb{P}$ in \eqref{eq::discritization::systemmatrix} contain singular integrals.
Therefore, we use the well-known Duffy trick, i.e., a regularizing coordinate transform for the quadrature of these singular integrals~\cite{Duffy_1982aa}.
Details on the implementation can be found in~\cite[App.~C.2]{Harbrecht_2001aa}.
For the non-singular integrals in these matrix blocks we use tensor product Gauss-Legendre quadrature.
The quadrature degree is increased logarithmically with the distance between the integrated elements~\cite{Harbrecht_2006aa}.
Note that the quadrature of these two matrices can be straightforwardly parallelized.

We refer to $\pmb{S}$ as \emph{generalized} incidence matrix because it extends the classical incidence matrix for a (directed) graph to the higher-order case.
Finally, for the right-hand side, we compute the entries of the excitation as follows
\begin{align}
    v_{\mathrm{ex},i} &= \int_\Gamma \left(\pmb{E}_{\mathrm i}\times \pmb{n}_{\pmb{x}}\right) \cdot \pmb{\nu}_i \dd\Gamma_{\pmb{x}}.
\end{align}

\noindent
It is worth noting that the proposed higher-order method can be interpreted as a variant of the PEEC method \cite{Ruehli_1974aa}.
They even coincide in the special case of lowest-order basis functions, i.e., $p=1$.
In this case, the circuit interpretation is maintained and a netlist can be extracted (even for NURBS geometries) \cite[Sec.~2.7.3]{Ruehli_2015aa}.
This is in agreement with earlier results which showed this analogy for the electrostatic case in~\cite{Torchio_2023aa}.

\subsection{Low-Frequency Stabilization}\label{sec::deflation}
When considering the low-frequency limit $\omega \rightarrow 0$ the formulation needs to maintain charge neutrality to guarantee a physically correct solution~\cite[Sec.~II]{Mautz_1984aa}.
The complexity of the problem increases with the number of disconnected components as discussed in~\cite{Taskinen_2006aa,Qian_2009aa}.
Here, we limit the discussion to the case of a single body.

In the formulation that discretizes the electric potential, the condition of charge neutrality in the case of a single body is equivalent to the potential being mean-free~\cite[Sec.~II.D]{Sharma_2022aa}.
To achieve this in the A-EFIE formulation the deflation approach can be utilized as proposed in~\cite{Qian_2009aa}.
This approach involves first performing a suitable scaling of the system matrix by the frequency and then applying a regularization to remove the null space of the system matrix of \eqref{eq::discritization::systemmatrix} which occurs in the low-frequency limit.

For the description of the null space in the case of higher-order basis functions, we can proceed analogously to~\cite[Sec.~7.2]{Steinbach_2008aa} and define
\begin{equation} \label{eq.aT}
    \pmb{a} = \bigg(\underbrace{0,\dots 0}_{N_{\pmb{j}}},\scalar{\varphi_1,1},\dots\scalar{\varphi_{N_{\varphi}},1}\bigg)\trans,
\end{equation}
with $\{\varphi_1,\ldots,\varphi_{N_\varphi}\}$ being the basis of $\mathbb{S}^2_{\pmb{p},\pmb{\Xi}}(\Gamma)$, which is a generalization of the lowest-order case to higher-order basis functions.
As proposed in~\cite{Qian_2009aa}, we scale the system matrix of~\eqref{eq::discritization::systemmatrix} by the frequency, denote the result by $\pmb{Z}$, and obtain the regularized system matrix given by
\begin{equation}
    \pmb{Z} - \gamma \pmb{a}\pmb{a}\trans,\label{eq::discritization::deflation}
\end{equation}
where $\gamma=\operatorname{trace}(\pmb{Z})/N$ and $\pmb{a}$ from \eqref{eq.aT}.
\section{Numerical Results} \label{sec:results}
This section describes the numerical experiments carried out to verify the proposed method.
First, an academic example is computed to analyse the convergence to a known analytical solution.
Furthermore, for this example, the condition number of the system matrix with and without deflation, as well as the resulting error in the E-field, are examined up to the low-frequency limit.
Additionally, we apply our presented method to an advanced problem, an antenna feed with significant curvature.
The computed solution is compared with well-known commercial software.

The implementation of the spline-based A-EFIE is carried out in the framework of Bembel~\cite{Dolz_2020ac}.
Thereby, the matrix assembly of \eqref{eq::discritization::systemmatrix} is parallelized and quadrature routines for singular integrals are reused.
Subsequently, the linear system is solved with a direct solver from the Eigen library~\cite{eigen3}.
To apply fast methods, e.g., $\mathcal{H}^2$-matrices for the isogeometric EFIE~\cite{Dolz_2019ad}, iterative solvers promise further speed-up.
However, this requires efficient preconditioners, which is for example thoroughly analyzed for the EFIE, see~\cite{Adrian_2021aa} for an overview of this topic.
An approach for the A-EFIE is presented in~\cite{Chen_2022aa}.
The analysis of iterative solvers and preconditioning is out of the scope of this paper.

\begin{figure}[t]
    \centering
    \input{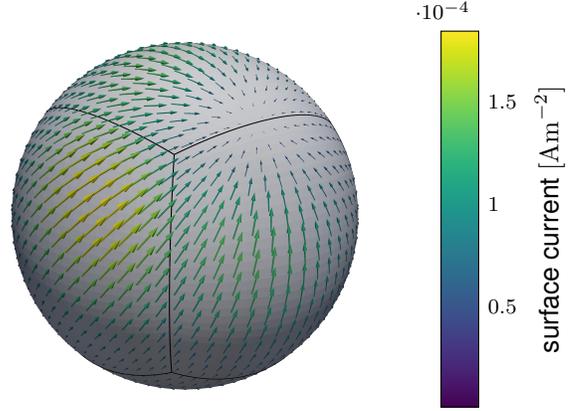}
    \caption{Visualization of the computed surface current excited by a dipole positioned inside the sphere discretized with six patches.}
\label{fig::dipole_splines}
\end{figure}
\subsection{Academic Example: Dipole}\label{subsec::results::dipole}
In the first numerical example, convergence of the spline-based A-EFIE method compared to a standard A-EFIE method is investigated.
For this purpose, the $E$-field of a dipole within a sphere with radius $\SI{1}{\meter}$ is used for excitation.
The surface quantities are computed, which then represent the $E$-field of the dipole in exterior of the sphere according to the equivalence theorem~\cite[Sec.~7.8]{Balanis_2016aa}.
This example is chosen because the analytical solution consists of two terms and can be evaluated easily.
Let the position ${\pmb{x}_0 = (0.2, 0.2, 0.2)}$ and orientation ${\pmb{p}_0 = (0, 0.1, 0.1)}$ of the dipole be given.
\mbox{The $E$-field of the dipole reads}
\begin{equation}
\begin{split}
    \pmb{E}_{\mathrm{DP}}(\pmb{x}) = \frac{e^{j\kappa r}}{4\pi\varepsilon}\biggl(&\frac{\kappa^2}{r}(\pmb{n}\times\pmb{p}_0)\times\pmb{n}\\
    &+\left(\frac{1}{r^3}-\frac{j\kappa}{r^2}\right)\left(3\pmb{n}(\pmb{n}\cdot\pmb{p}_0) - \pmb{p}_0\right)\biggl),\label{eq::numerics::dipole}
\end{split}
\end{equation}
with $r=|\pmb{x}-\pmb{x}_0|$, $\pmb{n}=(\pmb{x} - \pmb{x}_0)/r$, and $f=3$ MHz, cf.~\cite[Sec.~9.2]{Jackson_1998aa}.

In the spline-based A-EFIE method the sphere is represented by six NURBS patches with polynomial degree $p_1^n=p_2^n=4$, for $n=1,\dots6$, which is visualized in Fig.~\ref{fig::dipole_splines}.
Independently of the polynomial degree of the NURBS, we then select the polynomial degrees $p=1,2,3$ for the B-splines used to discretize the current, which is also shown in Fig.~\ref{fig::dipole_splines}.
Therefore, we refer to these results here and in the following examinations with the labels `B-splines $p=1,2,3$' respectively.
Results obtained with a standard A-EFIE formulation, which performs a lowest-order geometry approximation with triangles and uses linear polynomials to discretize the current, are labeled \mbox{`standard~A-EFIE'}.
\begin{figure}[t]
    \centering
    \begin{tikzpicture}
    \begin{customlegend}[legend columns=2,legend cell align=left,legend style={align=left,draw=black,font={\fontsize{10}{10}\sffamily}},legend entries={B-splines $p=1$, B-splines $p=2$, B-splines $p=3$, standard~A-EFIE}]

        \addlegendimage{Set1-A,line width=1pt,mark size=2pt,mark=*,every mark/.append style={fill=Set1-A!80!black}}
        \addlegendimage{Set1-B,line width=1pt,mark size=2pt,mark=square*,every mark/.append style={fill=Set1-B!80!black}}
        \addlegendimage{Set1-C,line width=1pt,mark size=2pt,mark=triangle*,every mark/.append style={fill=Set1-C!80!black}}
        \addlegendimage{Set1-D,line width=1pt,mark size=2pt,mark=halfsquare*,every mark/.append style={fill=Set1-D!80!black}}
    \end{customlegend}
\end{tikzpicture}

\begin{tikzpicture}
\pgfplotsset{
       every axis/.append style={
        font=\fontsize{10}{10}\sffamily},
      every non boxed x axis/.append style={
        x axis line style={->}
      },
      every non boxed y axis/.append style={
        y axis line style={->}
      },
      every non boxed z axis/.append style={
        z axis line style={->}
      },
      cycle list/Set1-4,
      cycle multiindex* list={
        mark list*\nextlist
        Set1-4\nextlist
      }, 
    }
\begin{loglogaxis}[
    height = 6cm,
    width = 0.48\textwidth,
    xlabel=number of degrees of freedom,
    ylabel=max. pw. error $E$-field,
    legend pos = south west,
    legend columns=1,
    legend cell align={left},
    grid = major,
    ymax=1,
    ymin=1e-12,
    xmin=10,
    xmax=2e4,
    ytick = {1,1e-3,1e-6,1e-9,1e-12},
    yticklabels = { 10\textsuperscript{0},
                    10\textsuperscript{-3},
                    10\textsuperscript{-6},
                    10\textsuperscript{-9},
                    10\textsuperscript{-12}},
    xtick = {1e1,1e2,1e3,1e4,1e5},
    xticklabels = { 10\textsuperscript{1},
                    10\textsuperscript{2},
                    10\textsuperscript{3},
                    10\textsuperscript{4}},
    every axis plot/.append style={line width=1pt,mark size=2pt},
    ]
    \addplot table [trim cells=true,x=all_dofs,y=error] {data/Dipole/P0.txt};
    \addplot table [trim cells=true,x=all_dofs,y=error] {data/Dipole/P1.txt};
    \addplot table [trim cells=true,x=all_dofs,y=error] {data/Dipole/P2.txt};
    \addplot table [trim cells=true,x=dofs,y=error] {data/Dipole/Ric.txt};
    \addplot+[black,dashed,no marks,samples at={20,10000}] {2^(-2*(log10(x/6)/log10(4)))*0.25}
        node[pos = 1,anchor=north, yshift=2pt] {$\mathcal{O}(h^2)$};
    \addplot+[black,dashed,no marks,samples at={20,10000}] {2^(-4*(log10(x/6)/log10(4)))*0.25}
        node[pos = 1,anchor=north, yshift=2pt] {$\mathcal{O}(h^4)$};
    \addplot+[black,dashed,no marks,samples at={20,10000}] {2^(-6*(log10(x/6)/log10(4)))*0.25}
        node[pos = 1,anchor=north, yshift=4pt] {$\mathcal{O}(h^6)$};

\end{loglogaxis}
\end{tikzpicture}
    \caption{Error for the example with the dipole inside the sphere.
    The maximum point wise error in the $E$-field computed with a (lowest-order) standard A-EFIE method and the spline-based A-EFIE method with increasing polynomial degree $p$ in discretizing $\pmb{j}$ is compared.
    }
    \label{fig::convergence_dipole}
\end{figure}
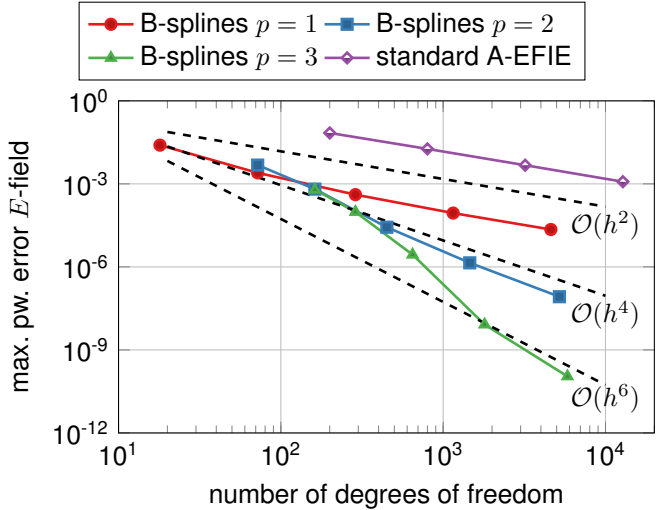
We compute the $E$-field in the exterior at a distance of $\SI{1}{m}$ from the sphere and compare point wise with \eqref{eq::numerics::dipole}.

We draw two conclusions from Fig.~\ref{fig::convergence_dipole}, which shows the error in the $E$-field against the number of degrees of freedom for the spline-based (of different polynomial degrees) and a standard A-EFIE.
First, even in the lowest-order case $p=1$ for the spline-based approach, the exact description of the geometry is advantageous compared to a standard A-EFIE.
While the rate of convergence is unchanged, accuracy per degree of freedom is improved.
Quantitatively speaking, consider a target accuracy of $10^{-3}$, then the spline-based approach requires only $N=288$, while a standard A-EFIE requires more than $N>10^4$ degrees of freedom.
Secondly, when increasing the polynomial degree of the B-spline ansatz functions representing the solution, the proposed spline-based A-EFIE shows higher-order convergence, \mbox{i.e., $\mathcal{O}(h^{2p})$} where $h$ is comparable with the mesh size.

Note, since there is no meshing step, the time to solution depends only on two contributions, matrix assembly and the effort to solve the resulting linear system of equations.
Even though quadrature of the spline-based method is more demanding compared to a method with a lowest-order geometry approximation, this is seamlessly parallelized.
Thus the significant bottle neck is the linear solver.
In this work, a direct solver is utilized, such that the relevant computational cost scales with $\mathcal{O}(N^3)$ and thus the accuracy per degree of freedom is a very relevant measure.
\begin{figure}[t]
    \centering
    \begin{tikzpicture}
    \begin{customlegend}[legend columns=2,legend cell align=left,legend style={align=left,draw=black,font={\fontsize{10}{10}\sffamily}},legend entries={B-splines $p=1$, B-splines $p=2$, B-splines $p=3$, standard~A-EFIE}]

        \addlegendimage{Set1-A,line width=1pt}
        \addlegendimage{Set1-B,line width=1pt}
        \addlegendimage{Set1-C,line width=1pt}
        \addlegendimage{Set1-D,line width=1pt}
    \end{customlegend}
\end{tikzpicture}

\begin{tikzpicture}
\pgfplotsset{
       every axis/.append style={
        font=\fontsize{10}{10}\sffamily},
      every non boxed x axis/.append style={
        x axis line style={->}
      },
      every non boxed y axis/.append style={
        y axis line style={->}
      },
      every non boxed z axis/.append style={
        z axis line style={->}
      },
      cycle list/Set1-4,
      cycle multiindex* list={
        mark list*\nextlist
        Set1-4\nextlist
      }, 
    }
\begin{loglogaxis}[
    height = 6cm,
    width = 0.48\textwidth,
    xlabel={frequency $[\si{\hertz}]$},
    ylabel={est.\ condition number},
    legend style={anchor=north east},
    legend columns = 1,
    legend cell align={left},
    grid = major,
    every axis plot/.append style={line width=1pt,mark size=2pt,no markers},
    ymin=1e1,
    ymax=1e28,
    xmin=1e-9,
    xmax=1e9,
    ytick = {1e1,1e10,1e19,1e28},
    yticklabels = { 10\textsuperscript{1},
                    10\textsuperscript{10},
                    10\textsuperscript{19},
                    10\textsuperscript{28}},
    xtick = {1e-9,
             1e-6,
             1e-3,
             1e0,
             1e3,
             1e6,
             1e9},
    xticklabels = { 10\textsuperscript{-9},
                    10\textsuperscript{-6},
                    10\textsuperscript{-3},
                    10\textsuperscript{0},
                    10\textsuperscript{3},
                    10\textsuperscript{6},
                    10\textsuperscript{9}},
    ]
    
    \addlegendimage{black,dashed}
    \addlegendentry{original}
    \addlegendimage{black}
    \addlegendentry{deflation}
    
    \addplot table [trim cells=true,x=f,y=condition] {data/Dipole/condition/P1M3_deflation.txt};
    \addplot table [trim cells=true,x=f,y=condition] {data/Dipole/condition/P2M3_deflation.txt};
    \addplot table [trim cells=true,x=f,y=condition] {data/Dipole/condition/P3M3_deflation.txt};
    
    \addplot table [trim cells=true,x=freq,y=cond] {data/Dipole/condition/ric_res_j999_phi666.txt};

    \addplot[Set1-A,dashed] table [trim cells=true,x=f,y=condition] {data/Dipole/condition/P1M3_original.txt};
    \addplot[Set1-B,dashed] table [trim cells=true,x=f,y=condition] {data/Dipole/condition/P2M3_original.txt};
    \addplot[Set1-C,dashed] table [trim cells=true,x=f,y=condition] {data/Dipole/condition/P3M3_original.txt};
    \addplot[Set1-D,dashed] table [trim cells=true,x=freq,y=cond] {data/Dipole/condition/ric_res_orig_sys_j999_phi666.txt};

\end{loglogaxis}
\end{tikzpicture}
    \vspace*{-1em}
    \caption{The estimated condition number of the original system from~\eqref{eq::discritization::systemmatrix} with B-splines with $p=1,2$ and $3$ is visualized over frequency with dashed lines.
    Accordingly, with solid lines showing the value for the matrix with applied deflation in~\eqref{eq::discritization::deflation}.
    In purple, the values of the respective matrices from the reference standard A-EFIE method.
    }
    \label{fig::condition_dipole}
\end{figure}
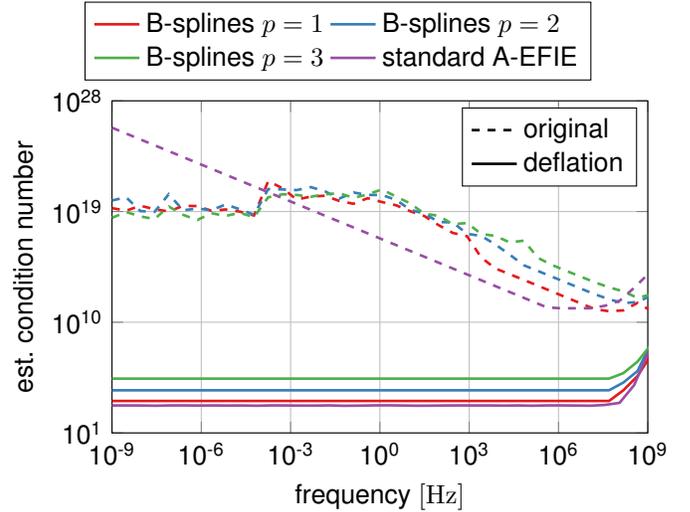
\begin{figure}[t]
    \centering
    \begin{tikzpicture}
    \begin{customlegend}[legend columns=2,legend cell align=left,legend style={align=left,draw=black,font={\fontsize{10}{10}\sffamily}},legend entries={B-splines $p=1$, B-splines $p=2$, B-splines $p=3$, standard~A-EFIE}]

        \addlegendimage{Set1-A,line width=1pt}
        \addlegendimage{Set1-B,line width=1pt}
        \addlegendimage{Set1-C,line width=1pt}
        \addlegendimage{Set1-D,line width=1pt}
    \end{customlegend}
\end{tikzpicture}

\begin{tikzpicture}
\pgfplotsset{
       every axis/.append style={
        font=\fontsize{10}{10}\sffamily},
      every non boxed x axis/.append style={
        x axis line style={->}
      },
      every non boxed y axis/.append style={
        y axis line style={->}
      },
      every non boxed z axis/.append style={
        z axis line style={->}
      },
      cycle list/Set1-4,
      cycle multiindex* list={
        mark list*\nextlist
        Set1-4\nextlist
      }, 
    }
\begin{loglogaxis}[
    height = 6cm,
    width = 0.48\textwidth,
    xlabel={frequency $[\si{\hertz}]$},
    ylabel={max. pw. error $E$-field},
    legend style={anchor=north east},
    legend columns = 1,
    legend cell align={left},
    grid = major,
    every axis plot/.append style={line width=1pt,mark size=2pt,no markers},
    ymin=1e-9,
    ymax=1e9,
    xmin=1e-9,
    xmax=1e9,
    ytick = {1e-9,1e-3,1e3,1e9},
    yticklabels = { 10\textsuperscript{-9},
                    10\textsuperscript{-3},
                    10\textsuperscript{3},
                    10\textsuperscript{9}},
    xtick = {1e-9,
             1e-6,
             1e-3,
             1e0,
             1e3,
             1e6,
             1e9},
    xticklabels = { 10\textsuperscript{-9},
                    10\textsuperscript{-6},
                    10\textsuperscript{-3},
                    10\textsuperscript{0},
                    10\textsuperscript{3},
                    10\textsuperscript{6},
                    10\textsuperscript{9}},
    ]
    
  \addlegendimage{black,dashed}
  \addlegendentry{original}
  \addlegendimage{black}
  \addlegendentry{deflation}
  
    \addplot table [trim cells=true,x=f,y=pw_error] {data/Dipole/condition/P1M3_deflation.txt};
    \addplot table [trim cells=true,x=f,y=pw_error] {data/Dipole/condition/P2M3_deflation.txt};
    \addplot table [trim cells=true,x=f,y=pw_error] {data/Dipole/condition/P3M3_deflation.txt};
    
    \addplot table [trim cells=true,x=freq,y=err] {data/Dipole/condition/ric_res_j999_phi666.txt};

    \addplot[Set1-A,dashed] table [trim cells=true,x=f,y=pw_error] {data/Dipole/condition/P1M3_original.txt};
    \addplot[Set1-B,dashed] table [trim cells=true,x=f,y=pw_error] {data/Dipole/condition/P2M3_original.txt};
    \addplot[Set1-C,dashed] table [trim cells=true,x=f,y=pw_error] {data/Dipole/condition/P3M3_original.txt};
    \addplot[Set1-D,dashed] table [trim cells=true,x=freq,y=err] {data/Dipole/condition/ric_res_orig_sys_j999_phi666.txt};

\end{loglogaxis}
\end{tikzpicture}
    \vspace*{-1em}
    \caption{The maximum point wise error of the $E$-field for the dipole example from Sec.~\ref{subsec::results::dipole} is visualized over frequency.
    With solid lines the error of the computation with the deflation method is indicated.
    Respectively, with dashed lines for the original method.
    }
    \label{fig::LF_error_dipole}
\end{figure}
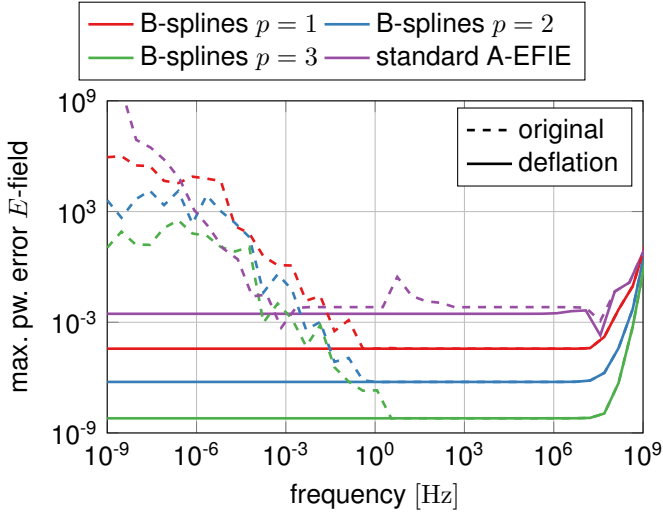

\begin{figure*}
    \centering
    \input{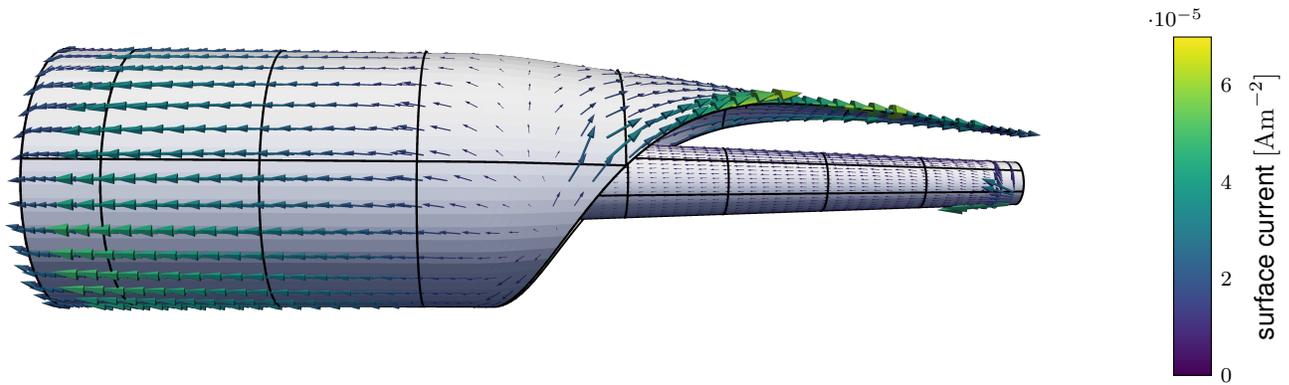}
    \vspace*{-1em}
    \caption{Representation of the coaxial balun geometry utilizing $82$ NURBS patches and visualization of the surface current discretized with B-splines.
    Both the inner and outer conductors are linearly tapered towards the spiral antenna, which is connected on the right, to match the impedance.
    In addition, the outer conductor is opened, which additionally leads to a mode transformation.
    See \cite{McParland_2022aa} for details of the dimensions.
    }
    \label{fig::balun-geometry}
\end{figure*}
\subsection{Low-Frequency Limit}
To investigate the proposed method in the low-frequency limit, we also use the dipole example from the previous section.
Firstly, the condition number of the system matrix with and without deflation is examined in dependency of the frequency.
Additionally, the error in the $E$-field is considered for the dipole reference solution from~\eqref{eq::numerics::dipole}.

A fixed discretization with polynomial degrees $p=1,2,3$ for the B-splines is investigated.
The knot vector is uniformly refined three times.
This results in $1152$, $1458$ and $1800$ degrees of freedom for the polynomial degrees $p=1,2$ and $3$, respectively.
In the computation with a standard A-EFIE $1665$ degrees of freedom are utilized.

The estimated condition numbers of the original system matrix of~\eqref{eq::discritization::systemmatrix}, i.e., without deflation, for both the spline-based and a standard A-EFIE are shown with dashed lines in Fig.~\ref{fig::condition_dipole}.
In combination with the considered error in the $E$-field, in Fig.~\ref{fig::LF_error_dipole} also shown with dashed lines, this shows the expected low-frequency breakdown of this formulation~\cite[Sec.~2.3]{Qian_2008aa}.
The application of the deflation approach~\eqref{eq::discritization::deflation}, i.e., solid lines in Fig.~\ref{fig::condition_dipole}, shows that the condition number remains stable up to the low-frequency limit.
The deflation method does not exert a negative influence on the high accuracy of the spline-based A-EFIE over the entire frequency range as illustrated in Fig.~\ref{fig::LF_error_dipole} with solid lines.
As to be expected, for a fixed mesh size, the condition number increases with higher polynomial degree $p$~\cite{Collier_2013aa}.
Applying deflation for a fixed $p$ and refining the mesh for the spline-based A-EFIE increases the condition number by the same order as a standard (lowest-order) A-EFIE.

\subsection{Multi-Tapered Coaxial Balun}
In the following, we describe how the proposed spline-based A-EFIE is applied to a realistic model of an antenna feed with spiral antenna, which is depicted in the upper part of Fig.~\ref{fig::balun_spiral_discretization}.
The design challenge is to maximize the radiated power of the antenna.
Due to their design, cables and antennas have different impedances, between which a seamless transformation by a feed with minimal reflections is required.
This can be achieved, for example, with a multi-tapered coaxial balun, depicted in Fig.~\ref{fig::balun-geometry}, as proposed in~\cite{McParland_2022aa}.

For our investigations, we create a model based on the design proposed by~\cite{McParland_2022aa} with NURBS.
Using the spline-based A-EFIE, we can directly proceed with the simulations at this point.
In contrast, for conventional methods a mesh generation is necessary.
Although the design with linear tapers may appear fairly elementary, the shape with significant curvature poses a challenge for the meshing.
Following the successful generation of a mesh, verification is conducted using two well-known commercial software packages CST Studio Suite~\cite{cst_2023aa} and Altair Feko~\cite{feko_2024aa}.
Furthermore, a comparison with a standard A-EFIE utilizing lowest-order geometry approximation with quadrilaterals is carried out.

For the design of the geometry, all values for the balun and the spiral antenna are taken from \cite{McParland_2022aa}, only the dielectric substrate is not taken into account.
The geometry is excited by a voltage source on the lower part of the coaxial cable, which is straightforwardly possible due to the formulation with electric potentials.
To compare the numerical solutions, the impedance $Z$ from $\SI{250}{\mega\hertz}$ up to $\SI{10}{\giga\hertz}$ is used to describe the characteristics of the balun including the spiral antenna.

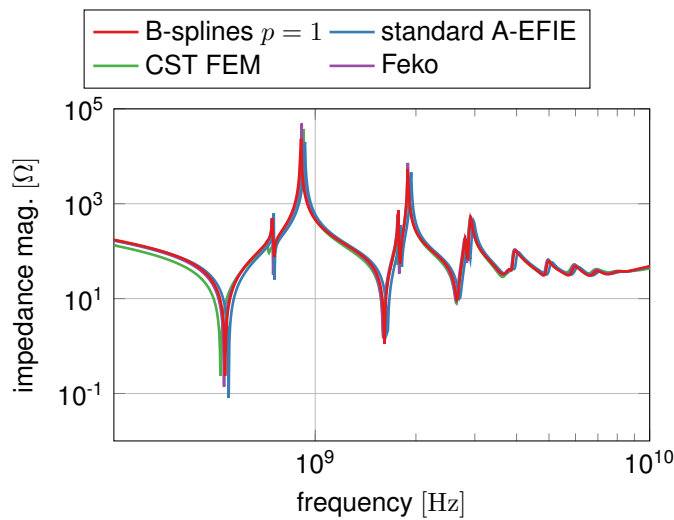
\begin{figure}
    \centering
    \begin{tikzpicture}
    \begin{customlegend}[legend columns=2,legend cell align=left,legend style={align=left,draw=black,font=\fontsize{10}{10}\sffamily},legend entries={B-splines~$p=1$, standard~A-EFIE, CST FEM, Feko}]

        \addlegendimage{Set1-A,line width=1pt}
        \addlegendimage{Set1-B,line width=1pt}
        \addlegendimage{Set1-C,line width=1pt}
        \addlegendimage{Set1-D,line width=1pt}
    \end{customlegend}
\end{tikzpicture}

\begin{tikzpicture}
\pgfplotsset{
       every axis/.append style={
        font=\fontsize{10}{10}\sffamily},
      every non boxed x axis/.append style={
        x axis line style={->}
      },
      every non boxed y axis/.append style={
        y axis line style={->}
      },
      every non boxed z axis/.append style={
        z axis line style={->}
      },
      cycle list/Set1-4,
    }

  \begin{axis}[
    xlabel={frequency $[\si{\hertz}]$},
    ylabel={impedance mag.\ $[\si{\ohm}]$},
    xmode=log,
    ymode=log,
    grid=major,
    width=0.48\textwidth,
    height=6cm,
    xmin=2.5e8,
    xmax=1e10,
    ymin=1e-2,
    ymax=1e5,
    reverse legend,
    ytick = {1e-1,1e1,1e3,1e5},
    yticklabels = { 10\textsuperscript{-1},
                    10\textsuperscript{1},
                    10\textsuperscript{3},
                    10\textsuperscript{5}},
    xtick = {1e9,1e10},
    xticklabels = { 10\textsuperscript{9},
                    10\textsuperscript{10}},
    legend style={at={(0.5,1.05)},anchor=south},
    legend columns = 2,
    every axis plot/.append style={line width=1pt,mark size=2pt,no markers},
  ]
  
  \addplot[Set1-C] table [x=frequency,
                  y=Zmag,
                  col sep=comma,
                  x expr={1e9*\thisrow{frequency}},
                  ] {data/Balun/cst_331k_500.txt};

  \addplot[Set1-D] table [x=frequency,
                  y=Zmag,
                  col sep=comma,
                  ] {data/Balun/feko_7213_1000.txt};

  \addplot[Set1-B] table [x=frequency,
                  y=Zmag,
                  col sep=comma,
                  ] {data/Balun/peec_8781_1000.txt};

  \addplot[Set1-A] table [x=frequency,
                  y=Zmag,
                  col sep=comma,
                  ] {data/Balun/bembel_1084_release.txt};
  \end{axis}
\end{tikzpicture}
    \vspace{-1em}
    \caption{Frequency sweep of the impedance computed to compare the numerical methods.
    The computations of the spline-based A-EFIE and the CST Studio Suite finite element method are carried out with $500$ frequency samples.
    For the computation carried out with standard A-EFIE and Feko $800$ samples are used.}
    \label{fig::Zparameter}
\end{figure}

The frequency sweeps of the impedance computed with four different methods/implementations are shown in Fig.~\ref{fig::Zparameter}.
The NURBS geometry, shown in the upper part of Fig.~\ref{fig::balun_spiral_discretization} consists of $N_\Gamma = 271$ patches and is computed with the spline-based A-EFIE on the first level of uniform refinement of the knot vector, i.e., with $1084$ elements.
The polynomial degree of the NURBS used here for the CAD geometry representation is at most $14$.
The B-spline basis functions for the discretization of the current are of polynomial degree $p=1$, i.e. this model allows a PEEC-like circuit extraction.
The frequency sweep of the impedance agrees very well with the solution obtained with integral equation method implemented by Feko.
However, the solution obtained with Feko used $7213$ elements.
In comparison with a standard A-EFIE, also with lowest-order geometry approximation utilizing $8781$ quadrilaterals, a similar curve of the impedance is obtained.
Finally, the solution is compared with a finite element method based on $331\cdot10^3$ tetrahedrons.
The model is excited by a discrete port in CST Studio Suite.
Again, a good agreement of the frequency sweep of the impedance is observed.
The small variations in the solutions are attributed to method and model differences, e.g., finite instead of boundary elements, geometry representations and meshing, as they are crucial when computing structures with significant curvature.

\section{Conclusion} \label{sec:conclusions}
This paper proposes a full-wave integral equation method based on the function spaces from isogeometric analysis.
A higher-order spline-based formulation of the augmented electric field integral equation has been proposed which avoids the low-frequency breakdown problem.
Thanks to the use of an integral equation formulation and the adoption of spline-based geometry concepts from isogeometric analysis the costly meshing step can be bypassed.
The presented spline-based A-EFIE method shows a higher accuracy per DOF compared to a standard A-EFIE method.
In the higher-order case, convergence rates of order~$2p$ have been observed for B-spline basis functions with polynomial degree~$p$.
Thus problems with complex geometries (e.g., that contain many curved parts) can be computed accurately with less degrees of freedom compared to a method utilizing lowest-order geometry approximations.

Future work will consider adaptive mesh refinement, e.g.,~\cite{Buffa_2022aa}, rigorous analysis of the convergence rate and model order reduction techniques for frequency sweeps.
The extension to dielectric media will also be considered~\cite{Xia_2016aa} and, finally, the proposed approach will be coupled with optimization tools for automatically optimizing complex shape structures.

\section*{ACKNOWLEDGEMENT}
This work was supported in part by the Graduate School CE within the Centre for Computational Engineering at Technische Universität Darmstadt, by the German Research Foundation under Project 443179833, the DAAD in the Framework of Short-Term under Grant 57588366 and the Future Talent Guest Stay program of TU Darmstadt.
JD acknowledges the support by the DFG under Germany’s Excellence
Strategy – project number 390685813.
The authors thank Stefan Kurz (Robert Bosch GmbH) and Frank Demming-Janssen (SIMUSERV GmbH) for the fruitful discussions on the topic.



\begin{thebibliography}{10}
\providecommand{\url}[1]{#1}
\csname url@samestyle\endcsname
\providecommand{\newblock}{\relax}
\providecommand{\bibinfo}[2]{#2}
\providecommand{\BIBentrySTDinterwordspacing}{\spaceskip=0pt\relax}
\providecommand{\BIBentryALTinterwordstretchfactor}{4}
\providecommand{\BIBentryALTinterwordspacing}{\spaceskip=\fontdimen2\font plus
\BIBentryALTinterwordstretchfactor\fontdimen3\font minus
  \fontdimen4\font\relax}
\providecommand{\BIBforeignlanguage}[2]{{%
\expandafter\ifx\csname l@#1\endcsname\relax
\typeout{** WARNING: IEEEtran.bst: No hyphenation pattern has been}%
\typeout{** loaded for the language `#1'. Using the pattern for}%
\typeout{** the default language instead.}%
\else
\language=\csname l@#1\endcsname
\fi
#2}}
\providecommand{\BIBdecl}{\relax}
\BIBdecl

\bibitem{Jin_2010aa_MoM}
J.-M. Jin, \emph{Theory and Computation of Electromagnetic Fields}.\hskip 1em
  plus 0.5em minus 0.4em\relax Hoboken, New Jersey: Wiley, 2010, p. 399.

\bibitem{Colton_2013aa_Ch9}
D.~Colton and R.~Kress, \emph{Integral Equation Methods in Scattering
  Theory}.\hskip 1em plus 0.5em minus 0.4em\relax Philadelphia: {Society for
  Industrial and Applied Mathematics ({SIAM})}, 2013, p. 244.

\bibitem{Adrian_2021aa}
S.~B. Adrian, A.~Dély, D.~Consoli, A.~Merlini, and F.~P. Andriulli,
  ``Electromagnetic integral equations: Insights in conditioning and
  preconditioning,'' \emph{{IEEE} Trans. Antenn. Propag.}, vol.~2, pp.
  1143--1174, 2021.

\bibitem{Wilton_1981aa}
D.~Wilton and A.~Glisson, ``On improving the stability of the electric field
  integral equation at low frequency,'' in \emph{Proceedings of {IEEE} Antennas
  and Propagation Society International Symposium}, 1981, pp. 124--133.

\bibitem{Mautz_1984aa}
J.~Mautz and R.~Harrington, ``An {E-Field} solution for a conducting surface
  small or comparable to the wavelength,'' \emph{{IEEE} Trans. Antenn.
  Propag.}, vol.~32, no.~4, pp. 330--339, 04 1984.

\bibitem{Zhao_2000aa}
J.-S. Zhao and W.~C. Chew, ``Integral equation solution of {Maxwell}'s
  equations from zero frequency to microwave frequencies,'' \emph{{IEEE} Trans.
  Antenn. Propag.}, vol.~48, no.~10, pp. 1635--1645, 10 2000.

\bibitem{Andriulli_2012aa}
F.~P. Andriulli, ``Loop-star and loop-tree decompositions: Analysis and
  efficient algorithms,'' \emph{{IEEE} Trans. Antenn. Propag.}, vol.~60, no.~5,
  pp. 2347--2356, 2012.

\bibitem{Bourhis_2024aa}
J.~Bourhis, A.~Merlini, and F.~P. Andriulli, ``High-order quasi-{Helmholtz}
  projectors: Definition, analyses, algorithms,'' \emph{{IEEE} Trans. Antenn.
  Propag.}, vol.~72, no.~4, pp. 3572--3579, 04 2024.

\bibitem{Taskinen_2006aa}
M.~Taskinen and P.~Yla-Oijala, ``Current and charge integral equation
  formulation,'' \emph{{IEEE} Trans. Antenn. Propag.}, vol.~54, no.~1, pp.
  58--67, 01 2006.

\bibitem{Qian_2010ab}
Z.-G. Qian and W.~C. Chew, ``Enhanced {A-EFIE} with perturbation method,''
  \emph{{IEEE} Trans. Antenn. Propag.}, vol.~58, no.~10, pp. 3256--3264, 10
  2010.

\bibitem{Qian_2008aa}
Z.~G. Qian and W.~C. Chew, ``\BIBforeignlanguage{english}{An augmented electric
  field integral equation for high-speed interconnect analysis},''
  \emph{\BIBforeignlanguage{english}{{IEEE} Trans. Antenn. Propag.}}, vol.~50,
  no.~10, pp. 2658--2662, 2008.

\bibitem{Qian_2009aa}
Z.-G. Qian and W.~C. Chew, ``Fast full-wave surface integral equation solver
  for multiscale structure modeling,'' \emph{{IEEE} Trans. Antenn. Propag.},
  vol.~57, no.~11, pp. 3594--3601, 11 2009.

\bibitem{Ruehli_1974aa}
A.~E. Ruehli, ``\BIBforeignlanguage{english}{Equivalent circuit models for
  three-dimensional multiconductor systems},''
  \emph{\BIBforeignlanguage{english}{{IEEE} Trans. Microw. Theor. Tech.}},
  vol.~22, no.~3, pp. 216--221, 1974.

\bibitem{Sharma_2022aa}
S.~Sharma and P.~Triverio, ``Electromagnetic modeling of lossy interconnects
  from {DC} to high frequencies with a potential-based boundary element
  formulation,'' \emph{{IEEE} Trans. Microw. Theor. Tech.}, vol.~70, no.~8, pp.
  3847--3861, 08 2022.

\bibitem{Xia_2016aa}
T.~Xia, H.~Gan, M.~Wei, W.~C. Chew, H.~Braunisch, Z.~Qian, K.~Aygün, and
  A.~Aydiner, ``An enhanced augmented electric-field integral equation
  formulation for dielectric objects,'' \emph{{IEEE} Trans. Antenn. Propag.},
  vol.~64, no.~6, pp. 2339--2347, 06 2016.

\bibitem{Zhang_2022ab}
L.~Zhang and M.~S. Tong, ``Low-frequency analysis of lossy interconnect
  structures based on two-region augmented volume-surface integral equations,''
  \emph{{IEEE} Trans. Antenn. Propag.}, vol.~70, no.~4, pp. 2863--2872, 2022.

\bibitem{Strang_2008aa}
G.~Strang and G.~Fix, \emph{An Analysis of the Finite Element Method},
  2nd~ed.\hskip 1em plus 0.5em minus 0.4em\relax Wellesley-Cambridge Press,
  2008.

\bibitem{Boggs_2005aa}
P.~T. Boggs, A.~Altshuler, A.~R. Larzelere, E.~J. Walsh, R.~L. Clay, and M.~F.
  Hardwick, ``\BIBforeignlanguage{english}{{DART} system analysis},'' Sandia
  National Laboratories, Technical Report SAND2005-4647, 08 2005.

\bibitem{Kang_2001aa}
G.~Kang, J.~Song, W.~C. Chew, K.~Donepudi, and J.-M. Jin, ``A novel grid-robust
  higher order vector basis function for the method of moments,'' \emph{{IEEE}
  Trans. Antenn. Propag.}, vol.~49, no.~6, pp. 908--915, 2001.

\bibitem{Graglia_1997aa}
R.~Graglia, D.~Wilton, and A.~Peterson, ``Higher order interpolatory vector
  bases for computational electromagnetics,'' \emph{{IEEE} Trans. Antenn.
  Propag.}, vol.~45, no.~3, pp. 329--342, 03 1997.

\bibitem{Jorgensen_2004aa}
E.~Jorgensen, J.~Volakis, P.~Meincke, and O.~Breinbjerg, ``Higher order
  hierarchical {Legendre} basis functions for electromagnetic modeling,''
  \emph{{IEEE} Trans. Antenn. Propag.}, vol.~52, no.~11, pp. 2985--2995, 11
  2004.

\bibitem{Ganesh_2008aa}
M.~Ganesh and S.~Hawkins, ``A high-order tangential basis algorithm for
  electromagnetic scattering by curved surfaces,'' \emph{{IEEE} Trans. Antenn.
  Propag.}, vol. 227, no.~9, pp. 4543--4562, 2008.

\bibitem{Weggler_2011aa}
L.~Weggler, ``\BIBforeignlanguage{english}{High order boundary element
  methods},'' Dissertation, Universität des Saarlandes, Saarbrücken, 08 2011.

\bibitem{Bruno_2020aa}
O.~P. Bruno and E.~Garza, ``A {Chebyshev}-based rectangular-polar integral
  solver for scattering by geometries described by non-overlapping patches,''
  \emph{{IEEE} Trans. Antenn. Propag.}, vol. 421, p. 109740, 2020.

\bibitem{Hellicar_2008aa}
A.~D. Hellicar, J.~S. Kot, G.~James, and G.~K. Cambrell, ``A comparison of
  higher order nodal- and edge-basis functions in the {MFIE} on rational
  {Bezier} geometries,'' \emph{{IEEE} Trans. Antenn. Propag.}, vol.~56, no.~6,
  pp. 1812--1818, 06 2008.

\bibitem{Hu_2021aa}
J.~Hu, E.~Garza, and C.~Sideris, ``A {Chebyshev}-based high-order-accurate
  integral equation solver for {Maxwell}’s equations,'' \emph{{IEEE} Trans.
  Antenn. Propag.}, vol.~69, no.~9, pp. 5790--5800, 2021.

\bibitem{Li_2016aa}
J.~Li, D.~Dault, B.~Liu, Y.~Tong, and B.~Shanker,
  ``\BIBforeignlanguage{english}{Subdivision based isogeometric analysis
  technique for electric field integral equations for simply connected
  structures},'' \emph{\BIBforeignlanguage{english}{J. Comput. Phys.}}, vol.
  319, pp. 145--162, 2016.

\bibitem{Simpson_2018aa}
R.~N. Simpson, Z.~Liu, R.~Vázquez, and J.~A. Evans,
  ``\BIBforeignlanguage{english}{An isogeometric boundary element method for
  electromagnetic scattering with compatible b-spline discretizations},''
  \emph{\BIBforeignlanguage{english}{J. Comput. Phys.}}, vol. 362, pp.
  264--289, 06 2018.

\bibitem{Dolz_2018aa}
J.~Dölz, H.~Harbrecht, S.~Kurz, S.~Schöps, and F.~Wolf,
  ``\BIBforeignlanguage{english}{A fast isogeometric {BEM} for the three
  dimensional {Laplace}- and {Helmholtz} problems},''
  \emph{\BIBforeignlanguage{english}{Comput. Meth. Appl. Mech. Eng.}}, vol.
  330, pp. 83--101, 03 2018, arxiv:1708.09162.

\bibitem{Dolz_2020aa}
J.~Dölz, S.~Kurz, S.~Schöps, and F.~Wolf, ``\BIBforeignlanguage{english}{A
  numerical comparison of an isogeometric and a parametric higher-order
  {Raviart}-{Thomas} approach to the electric field integral equation},''
  \emph{\BIBforeignlanguage{english}{{IEEE} Trans. Antenn. Propag.}}, vol.~68,
  no.~1, pp. 593--597, 01 2020, arxiv:1807.03628.

\bibitem{Fays_2023aa}
M.~Fays, O.~Chadebec, and B.~Ramdane, ``Isogeometric {FEM}-{BEM} coupling for
  magnetostatic problems modeling using magnetic scalar potential,''
  \emph{{IEEE} Trans. Magn.}, vol.~59, no.~5, pp. 1--4, 2023.

\bibitem{Simpson_2014ac}
R.~N. Simpson, M.~A. Scott, M.~Taus, D.~C. Thomas, and H.~Lian, ``Acoustic
  isogeometric boundary element analysis,'' \emph{Comput. Meth. Appl. Mech.
  Eng.}, vol. 269, pp. 265--290, 02 2014.

\bibitem{Chen_2020aa}
L.~L. Chen, Y.~Zhang, H.~Lian, E.~Atroshchenko, C.~Ding, and S.~P.~A. Bordas,
  ``Seamless integration of computer-aided geometric modeling and acoustic
  simulation: Isogeometric boundary element methods based on {Catmull-Clark}
  subdivision surfaces,'' \emph{Adv. Eng. Softw.}, vol. 149, p. 102879, 11
  2020.

\bibitem{Dolz_2019ad}
J.~Dölz, S.~Kurz, S.~Schöps, and F.~Wolf,
  ``\BIBforeignlanguage{english}{Isogeometric boundary elements in
  electromagnetism: Rigorous analysis, fast methods, and examples},''
  \emph{\BIBforeignlanguage{english}{{SIAM} J. Sci. Comput.}}, vol.~41, no.~5,
  pp. B983--B1010, 10 2019, arxiv:1807.03097.

\bibitem{Hofmann_2024aa}
B.~Hofmann, M.~Mirmohammadsadeghi, T.~F. Eibert, F.~P. Andriulli, and S.~B.
  Adrian, ``Low-frequency stabilization for the b-spline-based isogeometric
  discretization of the electric field integral equation,'' \emph{{IEEE} Trans.
  Antenn. Propag.}, vol.~72, no.~4, pp. 3558--3571, 04 2024.

\bibitem{Buffa_2019ac}
A.~Buffa, J.~Dölz, S.~Kurz, S.~Schöps, R.~Vázquez, and F.~Wolf,
  ``\BIBforeignlanguage{english}{Multipatch approximation of the de {Rham}
  sequence and its traces in isogeometric analysis},''
  \emph{\BIBforeignlanguage{english}{Numer. Math.}}, vol. 144, no.~1, pp.
  201--236, 06 2019, arxiv:1806.01062.

\bibitem{Dolz_2020ac}
J.~Dölz, H.~Harbrecht, S.~Kurz, M.~Multerer, S.~Schöps, and F.~Wolf,
  ``\BIBforeignlanguage{english}{Bembel: The fast isogeometric boundary element
  {C++} library for {Laplace}, {Helmholtz}, and electric wave equation},''
  \emph{\BIBforeignlanguage{english}{Software {X}}}, vol.~11, p. 100476, 04
  2020, arxiv:1906.00785.

\bibitem{Nolte_2024ab}
J.~Dölz, W.~Huang, M.~Multerer, M.~Nolte, R.~Von~Rickenbach, S.~Schöps, and
  F.~Wolf, ``Bembel: v1.1,'' 2024.

\bibitem{Torchio_2023aa}
R.~Torchio, M.~Nolte, S.~Schöps, and A.~E. Ruehli, ``A spline-based partial
  element equivalent circuit method for electrostatics,'' \emph{{IEEE} Trans.
  Dielectr. Electr. Insul.}, vol.~30, no.~2, 04 2023, arxiv:2207.13697.

\bibitem{Gope_2007aa}
D.~Gope, A.~Ruehli, and V.~Jandhyala, ``Solving low-frequency {EM-CKT} problems
  using the {PEEC} method,'' \emph{{IEEE} Trans. Adv. Packag.}, vol.~30, no.~2,
  pp. 313--320, 05 2007.

\bibitem{Hughes_2005aa}
T.~J.~R. Hughes, J.~A. Cottrell, and Y.~Bazilevs,
  ``\BIBforeignlanguage{english}{Isogeometric analysis: {CAD}, finite elements,
  {NURBS}, exact geometry and mesh refinement},''
  \emph{\BIBforeignlanguage{english}{Comput. Meth. Appl. Mech. Eng.}}, vol.
  194, pp. 4135--4195, 2005.

\bibitem{McParland_2022aa}
K.~P. McParland and M.~S. Mirotznik, ``Design and additive manufacture of
  multi-tapered coaxial baluns,'' \emph{{IEEE} Trans. Compon. Packag. Manuf.},
  vol.~12, no.~11, pp. 1806--1815, 11 2022.

\bibitem{Buffa_2003ab}
A.~Buffa and R.~Hiptmair, ``{Galerkin} boundary element methods for
  electromagnetic scattering,'' in \emph{Topics in computational wave
  propagation}, M.~Ainsworth, P.~Davies, D.~Duncan, B.~Rynne, and P.~Martin,
  Eds.\hskip 1em plus 0.5em minus 0.4em\relax Springer, 2003, pp. 83--124.

\bibitem{Stratton_1941aa}
J.~A. Stratton, \emph{\BIBforeignlanguage{english}{Electromagnetic
  Theory}}.\hskip 1em plus 0.5em minus 0.4em\relax {IEEE} Press, 1941.

\bibitem{Bendali_1984aa}
A.~Bendali, ``\BIBforeignlanguage{english}{Numerical analysis of the exterior
  boundary value problem for the time-harmonic {Maxwell} equations by a
  boundary finite element method. {I}. {The} continuous problem},''
  \emph{\BIBforeignlanguage{english}{Math. Comput.}}, vol.~43, no. 167, pp.
  29--46, 1984.

\bibitem{Buffa_2003aa}
A.~Buffa and S.~Christiansen, ``\BIBforeignlanguage{english}{The electric field
  integral equation on {Lipschitz} screens: definitions and numerical
  approximation},'' \emph{\BIBforeignlanguage{english}{Numer. Math.}}, vol.~94,
  no.~2, pp. 229--267, 2003.

\bibitem{Harrington_1968aa}
R.~F. Harrington, \emph{\BIBforeignlanguage{english}{Field Computation by
  Moment Methods}}.\hskip 1em plus 0.5em minus 0.4em\relax New York, US: The
  Macmillan Company, 1968.

\bibitem{Rao_1982aa}
S.~Rao, D.~Wilton, and A.~Glisson, ``Electromagnetic scattering by surfaces of
  arbitrary shape,'' \emph{{IEEE} Trans. Antenn. Propag.}, vol.~30, no.~3, pp.
  409--418, 05 1982.

\bibitem{Ruehli_2015aa}
A.~E. Ruehli, G.~Antonini, and L.~Jiang, \emph{\BIBforeignlanguage{english}{The
  Partial Element Equivalent Circuit Method for Electro-Magnetic and Circuit
  Problems}}, 1st~ed.\hskip 1em plus 0.5em minus 0.4em\relax John Wiley \&
  Sons, 2017.

\bibitem{Paul_2006aa}
C.~R. Paul, \emph{Introduction to Electromagnetic Compatibility}, 2nd~ed.\hskip
  1em plus 0.5em minus 0.4em\relax Hoboken, New Jersey: John Wiley \& Sons,
  2006.

\bibitem{Buffa_2002aa}
A.~Buffa, M.~Costabel, and D.~Sheen, ``\BIBforeignlanguage{english}{On traces
  for h(curl,{$\Omega$}) in lipschitz domains},''
  \emph{\BIBforeignlanguage{english}{J. Math. Anal. Appl.}}, vol. 276, no.~2,
  pp. 845--867, 2002.

\bibitem{Piegl_1997aa}
L.~Piegl and W.~Tiller, \emph{\BIBforeignlanguage{english}{The {NURBS} Book}},
  2nd~ed.\hskip 1em plus 0.5em minus 0.4em\relax Springer, 1997.

\bibitem{Beirao-da-Veiga_2014aa}
L.~Beirão~da Veiga, A.~Buffa, G.~Sangalli, and R.~Vázquez, ``Mathematical
  analysis of variational isogeometric methods,'' \emph{Acta. Num.}, vol.~23,
  pp. 157--287, 05 2014.

\bibitem{Peterson_2006aa}
A.~F. Peterson, ``\BIBforeignlanguage{english}{Mapped vector basis functions
  for electromagnetic integral equations},''
  \emph{\BIBforeignlanguage{english}{Synth. Lec. Comput. Electromagn.}},
  vol.~1, no.~1, pp. 1--124, 2006.

\bibitem{Duffy_1982aa}
M.~G. Duffy, ``\BIBforeignlanguage{english}{Quadrature over a pyramid or cube
  of integrands with a singularity at a vertex},''
  \emph{\BIBforeignlanguage{english}{{SIAM} J. Numer. Anal.}}, vol.~19, no.~6,
  pp. 1260--1262, 1982.

\bibitem{Harbrecht_2001aa}
H.~Harbrecht, ``\BIBforeignlanguage{english}{Wavelet {Galerkin} schemes for the
  boundary element method in three dimensions},'' Dissertation, Technische
  Universität Chemnitz, 2001.

\bibitem{Harbrecht_2006aa}
H.~Harbrecht and R.~Schneider, ``Wavelet {{Galerkin Schemes}} for {{Boundary
  Integral Equations---Implementation}} and {{Quadrature}},'' \emph{{SIAM} J.
  Sci. Comput.}, vol.~27, no.~4, pp. 1347--1370, 2006.

\bibitem{Steinbach_2008aa}
O.~Steinbach, \emph{\BIBforeignlanguage{english}{Numerical Approximation
  Methods for Elliptic Boundary Value Problems}}, ser. Finite and Boundary
  Elements.\hskip 1em plus 0.5em minus 0.4em\relax Springer, New York, 2008.

\bibitem{eigen3}
J.~Benoît and G.~Guennebaud, ``Eigen3 {C++} linear algebra template library,''
  official website, eigen.tuxfamily.org. Date of access July 17, 2024.

\bibitem{Chen_2022aa}
W.-J. Chen, S.~Sun, Y.~Liu, L.~Jiang, and J.~Hu, ``Improved {A-EFIE} system for
  electromagnetic simulation in low frequency regime,'' \emph{{IEEE} Antennas
  Wirel. Propag. Lett.}, vol.~21, no.~9, pp. 1752--1756, 09 2022.

\bibitem{Balanis_2016aa}
C.~A. Balanis, \emph{Antenna Theory: Analysis and Design}, 4th~ed.\hskip 1em
  plus 0.5em minus 0.4em\relax Hoboken, New Jersey: Wiley, 2016.

\bibitem{Jackson_1998aa}
J.~D. Jackson, \emph{\BIBforeignlanguage{english}{Classical Electrodynamics}},
  3rd~ed.\hskip 1em plus 0.5em minus 0.4em\relax New York, NY, USA: Wiley {\&}
  Sons, 1998.

\bibitem{Collier_2013aa}
N.~Collier, L.~Dalcin, D.~Pardo, and V.~M. Calo, ``The cost of continuity:
  Performance of iterative solvers on isogeometric finite elements,''
  \emph{{SIAM} J. Sci. Comput.}, vol.~35, no.~2, pp. A767--A784, 2013.

\bibitem{cst_2023aa}
\BIBentryALTinterwordspacing
{Dassault Systèmes}, ``{CST Studio Suite 2023},'' date of access December 20,
  2023. [Online]. Available:
  \url{https://www.3ds.com/products/simulia/cst-studio-suite}
\BIBentrySTDinterwordspacing

\bibitem{feko_2024aa}
\BIBentryALTinterwordspacing
{Altair}, ``{Feko},'' date of access June 4, 2024. [Online]. Available:
  \url{https://altair.com/feko}
\BIBentrySTDinterwordspacing

\bibitem{Buffa_2022aa}
A.~Buffa, G.~Gantner, C.~Giannelli, D.~Praetorius, and R.~Vázquez,
  ``\BIBforeignlanguage{english}{Mathematical foundations of adaptive
  isogeometric analysis},'' \emph{\BIBforeignlanguage{english}{Arch. Comput.
  Methods Eng.}}, vol.~29, no.~7, pp. 4479--4555, 11 2022.

\end{thebibliography}
\end{document}